# Aging and Domain Growth in Potassium-Lithium Tantalate Crystals

F. ALBERICI-KIOUS [(1)], J.-P. BOUCHAUD [(2)], L. F. CUGLIANDOLO [(3)],

P. DOUSSINEAU [(1)] and A. LEVELUT [(1)]

(1) *Laboratoire des Milieux Désordonnés et Hétérogènes* *
*Université P. et M. Curie, Case 86, 75252 Paris Cedex 05, France.*

(2) *Service de Physique de l'État Condensé*
*CEA-Saclay, 91191 Gif sur Yvette Cedex , France.*

(3) *Laboratoire de Physique Théorique, École Normale Supérieure* *
*24 rue Lhomond, 75231 Paris Cedex 05, France*
*and*
*Laboratoire de Physique Théorique et Hautes Énergies* *
*4 place Jussieu, 75252 Paris Cedex 05, France.*

# Abstract

We present new experimental results on the dielectric constant in orientationnal glasses $K_{1-x}Li_xTaO_3$ (KLT) with $x \leq 0.05$, together with the detailed (analytic and numerical) study of a model which attributes the observed aging to the motion of the walls of polarization domains. We show that the dielectric constant after a positive temperature jump goes through a maximum as a function of the subsequent time. This observation and those previously reported (aging, cooling rate dependences, etc) are compared with the predictions of the model, in which the variations of the dielectric constant are attributed to the change of polarization domain wall area. The total area decreases by domain growth and increases by nucleation of new small domains inside the large ones. These two opposite variations are both hindered by static random fields (equivalent to energy barriers) due to the frozen dipoles borne by the off-center $Li^+$ ions. Many results are well explained by a model with a single energy barrier. However, some effects can only be understood if a broad distribution of energy barriers is assumed. We use the experimental data to determine this distribution and find it to be unimodal with a width comparable to its most probable value.

# 1. Introduction

Ergodicity is a usual assumption of Statistical Mechanics. A system that satisfies this hypothesis is able to visit all allowable points of its phase space which are specified by the macroscopic constraints. In other words, the time average and the ensemble average of any thermodynamic quantity coincide. This general rule encounters some important exceptions: for instance, when a phase transition occurs, the associated symmetry breaking induces ergodicity breaking into mutually inaccessible regions of phase space [1].

Indeed, the dynamics and time scales play an important role. If the phase space is entirely visited in a finite time, the system is ergodic. If the phase space is splitted into several sub-spaces separated by barriers which diverge with the size of the system, the representative point of the instantaneous state may not be able to visit all space, even after times which diverge, thought not fast enough to allow it to surmount the diverging barriers. Then it is said that ergodicity is really broken (true ergodicity breaking). Another example is a case in which the arrival point (free energy minimum) in the phase space is separated of the departure point by an infinite number of finite barriers. Then the representative point will certainly arrive into the minimum, but this may take an infinite time (weak ergodicity breaking [2]).

From the experimentalist's point of view, things are not so clearly cut: actually, it may be practically impossible to distinguish between a finite time of one year and an infinite time. This somewhat fuzzy case (effective ergodicity breaking) corresponds to slow dynamics as observed in spin-glasses and many other systems.

The slow dynamics of orientational glasses (OG) has been extensively studied in the last years. The attention has been focused on two points: aging and ergodicity breaking. OG are materials which bear electric or elastic moments with orientational degrees of freedom distributed on random sites [3, 4]. In these crystals, each dipole or quadrupole is submitted to conflicting interactions from the ligand field and the fields generated by other moments located at random. Since the bearers of the moments are sticked to a given site, the disorder of their interactions is frozen in. This leads to frustration: the dipoles are not free to rotate; their moment, locked by high potential barriers, are not able to minimize the total free energy of the system.

Among other OGs, the potassium-lithium tantalate series $K_{1-x}Li_xTaO_3$ (KLT in short) has been particularly studied. The time-dependent response after different perturbations was measured. Examples of such response are: the birefringence after switch off of a mechanical stress [5], the elastic constant $c(\omega,t)$ [6] and, mainly, the dielectric constant $\varepsilon(\omega,t)$ after different thermal histories (regular temperature ramp, quenching, temperature jump, temperature cycle, ...) where $\omega$ is the circular frequency related to the frequency $f$ by $\omega = 2\pi f$ [7-9]. The alternative dielectric constant $\varepsilon(\omega,t)$ was found to slowly decrease with time, showing a non stationary behavior or aging. Using a fitting function, an extrapolated "asymptotic" value $\varepsilon(\omega,\infty)$ has been deduced. The dependence of $\varepsilon(\omega,\infty)$ on the cooling rate was discussed in terms of effective ergodicity breaking. In the end, a tremendous amount of data was thus gathered. But no microscopic model was proposed in order to explain the measurements.

In this paper we present a model which is intended to describe the whole set of data recorded on a series of KLT crystals. The model is based on the idea that the time dependent part of the dielectric constant (and to any other susceptibility such as the elastic constant) is due to the contribution of the wall of polarization domains, the size of which slowly evolves towards their equilibrium value. A short report has been recently published [10]. Concerning our experimental data, many of them are already published [6-9] and are only briefly recalled. Others are new and have a strong heuristic role for improving the model; therefore, they are reported in more detail.

The theoretical part of the paper is organized along two main lines: i) we first describe the scaling arguments, based on a domain growth picture, which lead to explicit expressions for the evolution of the dielectric constant for typical experiments; ii) we then perform a few numerical computations, on some particularly demonstrative experiments, which check or extend the validity of the previous calculations. In both cases, our attention is focused on the long time behavior.

We must emphasize that our model is fundamentally ergodic. Nevertheless, because of the time scales in play, it presents many of the features of ergodicity breaking: this is effective ergodicity breaking.

## 2. Experiments

**2.1. The Material** −  Pure $KTaO_3$ (KT) is a cubic perovskite crystal. It is called an incipient ferroelectric because its trend towards ferroelectricity is prevented by quantum fluctuations. Upon cooling, the displacements of the ions slow down but never become ordered. However, the correlation length of the polarization steadily increases as shown by the dielectric constant $\varepsilon$ which strongly rises as the temperature tends towards 0 K. The symmetry of the soft modes responsible for the enhancement of $\varepsilon$ corresponds to the displacements of the $Ta^{5+}$ ions along one of the six [100] directions [11, 12].

On random substitution of $Li^+$ ions to $K^+$ ions in a $KTaO_3$ crystal, the tendency towards ordering of the $Ta^{5+}$ ions is reduced [13]. Because of their small size, the $Li^+$ ions take off-center positions, referring to the centro-symmetric site. This generates electric dipoles at random sites which are aligned along one of the six [100] directions. Every dipole interacts with the others through the polarizable surrounding medium. At high temperatures, the dipoles constantly reorient between their six possible positions and the crystal is in a paraelectric phase. As temperature is lowered, the dipole reorientations slow down: the characteristic time for a jump over the barrier of $T_B \cong 1000$ K is longer than 3 years for $T < 24$ K. According to the $Li^+$ concentration $x$, the dipoles gradually freeze into a disordered phase ($x < 0.022$) or they suddenly freeze into a cooperative order ($x > 0.022$). Indeed, for any concentration $x$ larger than 0.01 there is a transition temperature $T_{tr}$ which separates the paraelectric phase from either the glassy or the ferroelectric phase. The phase obtained for low concentration is a frozen configuration with neither long-range ferroelectric order nor static polarization; it is an orientational or dipolar glass. We emphasize that the phase diagram thus described concerns the state of the $Li^+$ ions, not the state of the tantalate lattice that is always in a paraelectric phase. At low temperatures, its dielectric constant $\varepsilon$, while smaller than in pure KT, is rather large; this indicates the presence of large domains of fluctuating polarization.

**2.2. The Samples** − We have studied a series of mixed single crystals of formula $K_{1-x} Li_x TaO_3$. In addition to a nominally pure sample $(x < 0.001)$, we have examined five samples, the concentration $x$ and the transition temperature $T_{tr}$ of which are listed below: 0.011 (27 K); 0.017 (35 K); 0.025 (46 K); 0.030 (52 K); 0.050 (73 K). The lithium concentration $x$ is deduced according to the empirical law $T_{tr}(K) = 535\, x^{2/3}$ proposed by van der Klink *et al.* [13]. The error bars are $\Delta x = \pm 0.001$.

**2.3. The Measurement Methods** − All the faces of the samples are oriented perpendicular to a [100] direction of the cubic crystal. With chromium electrodes on two opposite end-faces, they form capacitors. The real and imaginary parts of the dielectric constant $\varepsilon(\omega) = \varepsilon'(\omega) - i\, \varepsilon''(\omega)$ are related to the capacitance and loss. They were measured with a Hewlett-Packard 4192A impedance analyser at seven frequencies ranging from 1 kHz to 1 MHz. Some elastic constant measurements were also performed, in the 100 MHz range. They provide the real and imaginary parts of the elastic constant $c(\omega) = c'(\omega) - i\, c''(\omega)$. In KLT the imaginary parts are much smaller than the real part and they are not used in what follows. It was checked that the response is linear. All the measurements were performed without any applied static electric field or mechanical stress, *i.e.*, without any symmetry breaking. Since the capacitance $C$ and the real part $\varepsilon'$ of the dielectric constant are simply proportional through a geometrical factor, our data are generally displayed in pF in the figures as they were recorded.

**2.4. Results Already Known** − Our experiments can be classified in several types according to the procedure used to attain the experiment temperature $T_{exp}$.

(i) Variable cooling ramp. − The sample is regularly cooled from a high temperature through the transition temperature $T_{tr}$ down to $T_{exp}$ at a constant rate $R = dT/dt$. Then, the dielectric constant $\varepsilon'(\omega, t)$ starts from $\varepsilon'(\omega, 0)$ and steadily decreases. This aging is very well fitted by a power law function $\varepsilon'(\omega, t) = \varepsilon(\omega, \infty) + \Delta\varepsilon\, ((t + t_0)/t_0)^{-\alpha}$ for times as long as $400\,000$ s. However, the main result of this analysis is that the apparent "asymptotic" value $\varepsilon(\omega, \infty)$ depends on the cooling rate with a larger $|R|$ leading to a higher $\varepsilon(\omega,\infty)$. More precisely, the dependence is approximately $\varepsilon(\omega,\infty) = p + q \log|R|$ (see figures 5 and 6 in reference [8]). All the samples share this behavior with a positive $q$, except for the nominally pure sample for which $q$ is slightly negative.

(ii) Interrupted cooling ramp. − The sample is cooled at a constant rate $R$ down to the quenching temperature $T_q$ below $T_{tr}$, from where it is very rapidly cooled at the rapid rate $R_q \cong 15\,R$ down to $T_{exp}$. The same $R$ and $R_q$ are used for a series of quenching temperatures with $T_{exp} < T_q < T_{tr}$. Figure 11 in reference [8] shows the dependence of the initial value $\varepsilon(\omega,0)$, at the beginning of the isothermal evolution at $T_{exp}$ together with the "asymptotic" value $\varepsilon(\omega,\infty)$ for the sample with $x = 0.025$. They both have the same behavior: they depend on $T_q$ if it is close to $T_{tr}$, but they become independent of $T_q$ if $T_q$ is small enough. The comparison of our samples with $x = 0.025$ and $x = 0.011$ shows that their behaviors are correlated with the dependences on the cooling rates: the larger coefficient $q$ in a cooling ramp experiment corresponds to the larger curvature in a quenching experiment. Therefore, these two results demonstrate that what is important for the further evolution of $\varepsilon'(\omega,t)$ is the manner in which the transition temperature is crossed.

(iii) Temperature cycles. − Two temperatures $T_1$ and $T_2$ are used in three steps. After the sample was regularly cooled down to $T_1$, it is kept in isothermal conditions during $t_1$. Then, the temperature is suddenly changed to $T_2$ and kept constant for $t_2$. Finally, the sample is suddenly driven back to $T_1$. Measurements are continuously taken all along the three periods. The temperature $T_2$ may be smaller or larger than the temperature $T_1$. If $T_2 > T_1$, the sojourn at $T_2$ acts as a simple annealling [8]. If $T_2 < T_1$, the experiment is much more interesting since a paradox appears: on the one hand, the system ages at $T_2$ during the time $t_2$; on the other hand, after the return to $T_1$, a continuous curve joining the first and third periods is obtained by simply translating the third period, just as if the second period had not existed (see figure 14 in reference [8]). Remarkable transient effects were observed after positive temperature jumps. They are reported in detail in the next sub-section.

Similar experiments have been performed in spin-glasses (for a review, see [14]).

**2.5. New Results** – During the course of temperature cycle experiments we had observed a surprising behavior which immediately followed a positive temperature jump. At the jump the dielectric constant starts from a high value at the low temperature $T_1$ and rapidly decreases to a smaller value at $T_2 > T_1$. But $\varepsilon'$ first reaches a low value, then it immediately increases back, passes by a maximum and finally resumes its slow downward aging [9]. We present here a more complete report on this phenomenon studied in the sample with $x = 0.017$.

At fixed low temperature $T_1$ and for a fixed duration $t_1$, we have performed jumps to different high temperature $T_2 > T_1$. The larger the temperature jump $T_2 - T_1$, the larger the downward jump of $\varepsilon'$. The behavior of $\varepsilon'(\omega, t)$ is always non monotonic (see figure 1), whatever $T_2$, lower or higher than the transition temperature $T_{tr}$. In this last case the final decrease is very small and rapid.

At fixed temperatures $T_1$ and $T_2$, we have studied the role of the duration $t_1$ which was varied between $500$ s and $80000$ s. The curve corresponding to the shortest $t_1$ is the steepest and it encompasses all the other. For the longest $t_1$, the curve has a flat maximum, lower than all the other maxima. When $t_1$ increases, the location of the maximum is shifted towards larger times. For long times, all curves merge into a common curve (see figure 2). We have analysed the curves as follows. They are decomposed in two parts: a long time decrease, shared by all the curves and well described by a single power law and a short time increase. The difference between the experimental data and the best power law separates the initial increase that may be described as an exponential return $\Delta\varepsilon_{incr}\left[1 - \exp(-t/\tau_{incr})\right]$. The amplitude $\Delta\varepsilon_{incr}$ rises with $t_1$ but saturates for large $t_1$. Surprisingly, $\Delta\varepsilon_{incr}$ is smaller for larger $T_2$ (see figure 3). The characteristic time $\tau_{incr}$ is nearly independent of $t_1$ (see figure 4) and it is longer for lower temperature $T_2$, approximately following the Arrhenius law with an activation barrier energy $T_b$ of the order of 15 K. Its value $\tau_{incr} \approx 1200$ s is short in comparison to that of aging in long experiments (of the order of $t_1$).

**2.6. Dielectric Measurements and Elastic Measurements** – The results obtained by dielectric measurements and elastic measurements show unexpected similarities. After a regular temperature ramp, the real part of the dielectric constant $\varepsilon'(\omega, t)$ and the real part $c'(\omega, t)$ of the elastic constant both isothermally decrease as time flows. After a positive temperature jump, both $\varepsilon'(\omega,t)$ and $c'(\omega,t)$ first increase, then pass by a maximum and finally decrease [6, 8]. This common behavior is surprising since it is well known that the interaction of a wave with relaxing entities always slows down the wave. This is true for the electromagnetic wave velocity $v_{em}$ as well as for the elastic wave velocity $v_{el}$. Therefore, since $v_{em} = c_l / \sqrt{\varepsilon' \mu}$ and $v_{el} = \sqrt{c'/\rho}$ ($c_l$ is the light velocity in vacuum, $\mu$ is the magnetic permittivity and $\rho$ is the specific mass) the dielectric constant $\varepsilon'$ and the elastic constant $c'$ are expected to present opposite behaviors. This is actually observed in KLT in the temperature range 40-60 K where $\varepsilon'$ increases while $c'$ decreases, due to the relaxation interaction with the Li$^+$ ions jumping over their potential barriers. Therefore, the two quantities which must be compared are the dielectric constant $\varepsilon'$ and the elastic compliance $s' = 1/c'$, inverse of the elastic constant or stiffness (the tensorial character of these quantities is neglected here). In this context, the similar behaviors observed on $\varepsilon'$ and $c'$ during aging were not expected at all.

The preceding remarks rule out, as source of aging, any relaxation process such as the thermally activated jumps of the Li$^+$ ions between equivalent off-center positions. They prompt to look for new interaction processes such that $\varepsilon'$ and $c'$ must behave in the same manner.

## 3. The Model

We recall here some important features concerning KLT which are relevant for the model. Firstly, the dielectric constant is large and increases as temperature is lowered, thus indicating the presence of electric polarization domains of the host lattice, oriented along one of its six [100] directions, with growing size. Secondly, below 40 K the Li$^+$ ions are completely frozen in and the dipoles they bear, also oriented along one of the six [100] directions of the lattice, generate electric static random fields. Thirdly, the highly polarizable host crystal is locally sensitive to the presence of the Li$^+$ ions: around any of them there is a distortion of the lattice which gives it a large effective dipolar moment, as measured in [15] and explained in [16].

Figure 5 shows the results of a numerical simulation of a three-dimensional assembly of Ising spins with ferromagnetic interactions in a random magnetic field. We present a two-dimensional cross-section of the system at four different times. The up-spins are represented by black dots and the down-spins by white dots. These pictures provide some information on the geometry of the domains of ordered spins. Indeed, even if their shapes are very whimsical, some features are clearly seen: they have a typical size which increases with time and there is also a visible size dispersion.

These facts are in favor of the role of the polarization domains in the time-dependent behavior of the dielectric constant $\varepsilon'$. Since $\varepsilon'(t)$ decreases when the typical domain size $r(t)$ increases (because the volume fraction of the walls then decreases), we will assume in the following that the major contribution to the aging part of the dielectric constant comes from the domain walls, and not from the bulk.

Actually, the pictures of figure 5, shown as an illustrative example for the domain shape, may be misleading and we have to emphasize the physical differences between those Ising spin domains and the polarization domains in KLT. On the one hand, the Ising spin domains which bear a quasi static magnetization are dipolar domains. On the other hand, the KLT lattice is in a paraelectric phase of an incipient ferroelectric. This means that there are polar soft modes, the frequency of which lowers when the temperature decreases, as evidenced by the increase of $\varepsilon'$. These domains are the forerunners of the aborted ferroelectric transition at 0 K. At any time they bear an electric polarization which oscillates at the soft mode frequency. Hence, the mean polarization taken over time intervals longer than $10^{-11}$ s (as for our experiments), vanishes. Consequently, for our purpose the KLT domains are rather quadrupolar than dipolar, the two directions along any of the three cubic axes cannot be distinguished and, finally, there are only three types of domains in KLT, oriented along the three [100] axes. An orientable entity in KLT passes from a domain to a neighboring one by a $\pi/2$-flip while for Ising domains this needs a $\pi$-flip.

**3.1. First Insight on the Model** − We assume that the time-dependent contribution to the complex dielectric constant $\varepsilon$ (and to the complex elastic constant $c$ too) is due to the polarization domain walls. These walls change in two manners. They have an oscillating motion, forced by the applied alternative electric field (or elastic strain); the motion is not in phase with the field and the subsequent hysteresis induces loss and dispersion. Moreover, if the domains are out of equilibrium, the walls also have a natural slow motion towards the equilibrium size, defined as the correlation length $\xi(T)$ of the fluctuations of polarization. The length $\xi(T)$ depends on temperature; its square is proportional to the equilibrium value of the dielectric constant, and consequently, $\xi(T) = \xi_{tr}\sqrt{T_{tr}/T}$ where the correlation length $\xi_{tr} = \xi(T_{tr})$ at the transition temperature $T_{tr}$ is taken as reference. The natural motion of the walls is hindered by the static random fields generated by the Li$^+$ ions when they are frozen (for $T < T_{tr}$); the motion is free when the Li$^+$ ions are permanently re-orientating (for $T > T_{tr}$), producing zero field in average. As a result of the two movements, the complex susceptibilities measured at the circular frequency $\omega$ also depend on time and read $\varepsilon(\omega,t)$ and $c(\omega,t)$. We notice that the oscillating motion of the domain walls corresponds to polarizability and elasticity added in parallel to those of the lattice, as discussed in sub-section 2.6. Because most experiments were dielectric measurements, in what follows we focus our attention on $\varepsilon(\omega,t)$ only.

Moreover, we assume that the loss and dispersion are proportional to the total wall area. The proportionality factor is an efficiency coefficient which contains the cosine of the angle between the field and the normal direction of the wall at any point of the surface.

As a first step, we assume that the size dispersion is weak enough to be neglected. Then all the domains have the same size $r(t)$. Then the area of the surface of a domain is $\sigma(t) \approx r(t)^2$ (a numeric coefficient of geometric origin is omitted), while the number of domains in the sample of volume $V$ is $n(t) \approx V/r(t)^3$ (here too, a numeric coefficient is omitted). Finally, it is found that the total wall area is $\Sigma(t) = C\,V/r(t)$, where the coefficient $C$ is a dimensionless geometrical factor which takes into account the orientation of the field and a factor of $1/2$ which prevents to count twice the surfaces which are shared by contiguous domains. In most experiments, isothermal aging is studied after cooling. Then the size $r(t)$ at temperature $T$ is smaller than $\xi(T) = \xi_{tr}\sqrt{T_{tr}/T}$ and the trend of the domains towards equilibrium can only be growth. If $r(t)$ increases, then $\Sigma(t)$ and $\varepsilon'(\omega,t)$ decrease, as actually observed.

Pinning impurities strongly impede the domain wall motion and lead to thermally activated motion [17]. We model this by writing a simple rate equation for the isothermal evolution of their size $\rho(t)$:

$$\frac{d\rho}{dt} \approx \frac{a}{\tau(\rho,T)} \qquad (1)$$

where

$$\tau(\rho,T) = \tau_\infty \exp\left(\frac{(\rho/a)^\theta T_\sigma}{T}\right).$$

$\theta$ is an exponent describing how the energy barriers typically grow with the length scale $\rho$ involved in the motion, $T_\sigma$ measures the typical strength of the pinning field while $a$ is of the order of the lattice cell parameter. The exponent $\theta$ depends on the coupling between the polar entities: $\theta \cong 4/3$ for nearest neighbor interactions while $\theta \cong 2$ for long range dipolar interactions. In the following we use $\theta = 1$, a value not far from the previous realistic ones, which allows equation (1) to be solved exactly as:

$$\rho(t) = \rho_0 + a(T/T_\sigma)\ln\left[1 + t/\tau(\rho_0,T)\right],$$

where $\rho_0$ is the initial value of the domain size. This function increases rapidly as $\rho(t) \approx \rho_0 + a(T/T_\sigma)t/\tau(\rho_0,T)$ for $t/\tau(\rho_0,T) \ll 1$ and slowly as $\rho(t) \approx a(T/T_\sigma)\ln\left[t/\tau(\rho_0,T)\right]$ for $t/\tau(\rho_0,T) \gg 1$. Accordingly, the free size $\rho(t)$ can go to infinity. In fact, the true size $r(t)$ is necessarily limited by the thermal correlation length $\xi(T)$, then

$$r(t) = \min\left\{\rho(t), \xi(T)\right\},$$

and the corresponding relaxation time becomes

$$\tau(r_i,T) = \tau_\infty \exp\left(\frac{r_i T_\sigma}{aT}\right) \quad \text{with} \quad r_i = r(t_i).$$

The relaxation time has the Arrhenius form with the barrier height $T_\sigma$, measured in Kelvins; $\tau_\infty$ is a microscopic time. This form means that the activation process responsible for the evolution for $T < T_{tr}$ is limited by the effective barrier height $\tilde{T}_\sigma(r_1) = (r_1/a)T_\sigma$ which depends on the initial domain size $r_1$ scaled by the lattice cell parameter $a$. If $T > T_{tr}$ the Li$^+$ ions are not frozen, then $T_\sigma = 0$ and $\tau(r_i,T) = \tau_\infty$ is very short.

The simplest experiment begins with a controlled cooling at constant cooling rate $R = dT/dt$ from the initial temperature $T_{init}$ above $T_{tr}$ to the experiment temperature $T_{exp}$ below $T_{tr}$. The temperature variation reads $T(t) = T_{tr} + R\,t$ with $R < 0$. The relevant cooling time is $t_c = (T_{tr} - T_{exp})/|R|$. The correlation length depends on time through $T(t)$. It reads $\xi(t) = \xi_{tr}\sqrt{T_{tr}/(T_{tr} + R\,t)}$ for $0 \leq t \leq t_c$ (or $T_{exp} \leq T \leq T_{tr}$) if the time origin is taken at the instant when $T_{tr}$ is crossed. The evolution velocity of $\xi(t)$ and the domain growth velocity respectively read

$$\frac{d\xi(T(t))}{dt} = \dot{\xi} = -\frac{R}{2}\frac{\xi(T(t))}{T(t)} \quad \text{and} \quad \frac{d\rho(t)}{dt} = \dot{\rho} = \frac{a}{\tau(\rho,T(t))}\ . \qquad (2)$$

As long as $\dot{\rho} > \dot{\xi}$ the domains are able to reach their maximum size $\xi(T)$ at any time during cooling. This is no longer true for $t > t^*$ where the time $t^*$ is determined by the condition $\dot{\rho}(t^*) = \dot{\xi}(T(t^*))$. Two cases have to be considered.

If $t^* > t_c$, during all the cooling and afterwards, the domains are in equilibrium. The domain size is

$$r(t) = \xi(T(t)) \qquad \text{for } 0 < t < t_c\ ,$$
$$r(t) = \xi(T_{exp}) \qquad \text{for } t > t_c\ .$$

Consequently, there is no isothermal aging since $r(t)$ does not vary for $t > t_c$.

If $0 < t^* < t_c$, the domains are in equilibrium up to $t^*$ and fall out of equilibrium when $t > t^*$. Afterwards, they non isothermally grow from $T^* = T(t^*)$, where $r(t^*) = \xi\left(T_{tr} + R\,t^*\right)$, down to $T_{exp}$ and finally, isothermally grow. Therefore, the domain size is

$$r(t) = \xi(T(t)) \qquad \text{for } 0 < t < t^*\ ,$$

$$r(t) = r(t^*) + \int_{t^*}^{t} dt'\,\frac{d\rho(t',T(t'))}{dt'} \qquad \text{for } t^* < t < t_c\ , \quad (3)$$

$$r(t) = \min\left\{r_c + a\,(T_{exp}/T_\sigma)\ln\left[1 + (t - t_c)/\tau(r_c, T_{exp})\right],\ \xi(T_{exp})\right\} \qquad \text{for } t > t_c\ ,$$

with $r_c = r(t_c)$.

The values of $R$ and $T_\sigma$ determine the case to be considered. Indeed, the time $t^*$ is the solution of the equation

$$\frac{d\xi(t)}{dt}\bigg)_{t^*} = \frac{d\rho(t)}{dt}\bigg)_{t^*}$$

which reads, with $T^* = T_{tr} + R\,t^*$,

$$-\frac{R}{2}\frac{\xi(T^*)}{T^*} = a\frac{(T^*/T_\sigma)}{\tau\big(\xi(T^*),T^*\big)} .$$

In this equation the right-hand side is the derivative of $\rho(t)$ written with $t = t^*$ and $r(t^*) = \xi(T^*)$ as initial values. After a trite calculation we find

$$T^* = \left( \frac{\xi_{tr}}{a} \frac{T_{tr}^{1/2}\,T_\sigma}{\ln\left( \dfrac{2}{|R|\,\tau_\infty} \dfrac{a}{\xi_{tr}} \dfrac{(T^*)^{5/2}}{T_{tr}^{1/2}\,T_\sigma} \right)} \right)^{2/3} . \qquad (4)$$

The logarithm is a slowly varying function if its argument is much larger than 1. Therefore the unknown temperature $T^*$ may be replaced in the right-hand side by any temperature between $T_{tr}$ and $T_{exp}$ without noticeable change. We arbitrarily choose $T^* = T_{tr}$ and set

$$R_0 = \frac{2}{\tau_\infty} \frac{a}{\xi_{tr}} \frac{T_{tr}^{\,2}}{T_\sigma}$$

to find, as a good approximation,

$$T^* \cong \left( \frac{\xi_{tr}}{a} \frac{T_{tr}^{1/2}\,T_\sigma}{\ln(R_0/|R|)} \right)^{2/3} .$$

With $\tau_\infty = 10^{-10}$ s, $R = -0.1$ K.s$^{-1}$ and $T_{tr} = 35$ K we obtain $T^* \cong 4.7$ K if $T_\sigma\,\xi_{tr}/a = 50$ K or $T^* \cong 37$ K if $T_\sigma\,\xi_{tr}/a = 1000$ K, while an order of magnitude for $R_0$ is $R_0 \approx 10^{11}$ K.s$^{-1}$, which is extremely large. Therefore, in the following, we shall often assume that $\ln R_0 \gg \ln|R|$. In the first case $T^* < T_{exp} = 5$ K and the domains are always in equilibrium: the characteristic time $\tau$ is still of the order of the microsecond at the lowest temperature $T_{exp} = 5$ K. In the second case $T^* > T_{tr}$, the domains fall out of equilibrium as soon as $T_{tr}$ is crossed (above $T_{tr}$ they have the equilibrium size $\xi(T)$): the characteristic time is already $\tau \approx 30000$ s at $T_{tr}$ and longer than the age of the universe at $T_{exp}$. In both cases there is no aging at $T_{exp}$, either because the domains are too rapidly in equilibrium or because they are completely frozen. Therefore, aging must be looked for in the intermediate range.

If $T^*$ is close to $T_{exp}$ the difference between the coherence length and the domain size can be expanded in powers of $(T^* - T_{exp})$ and to second order it reads

$$\xi(t_c) - \rho(t_c) = \xi_{tr} \frac{\left[T^* - T_{exp}\right]^2}{8 T_{tr}^{1/3} T_\sigma^{5/3} \left(\frac{\xi_{tr}}{a}\right)^{5/3}} \left(\ln(R_0/|R|)\right)^{5/3} \left[5 + 3 \ln(R_0/|R|)\right] \quad (5)$$

with $\xi(t_c) = \xi(T_{exp})$.

Writing $\ln(R_0/|R|) = \ln(R_0)\left[1 - \ln(|R|)/\ln(R_0)\right]$ and using $\ln(|R|)/\ln(R_0) \ll 1$, it is clear that the result may indeed be put in the form $\xi(T_{exp}) - \rho(t_c) \cong A + B \ln(|R|)/\ln(R_0)$.

The corresponding variation of $\varepsilon'$ is

$$\varepsilon'(t_c) - \varepsilon'(\infty) \cong C + D \ln(|R|)/\ln(R_0) \quad . \tag{6}$$

as experimentally observed [8].

Another simple experiment is quenching. It begins at the temperature $T_{tr}$, with a controlled cooling at constant rate $R = dT/dt$, interrupted at the temperature $T_q$ by a sudden (instantaneous, in principle) jump down to $T_{exp}$. The quenching happens at $t_q = (T_{tr} - T_q)/|R|$. The correlation length is $\xi(t) = \xi_{tr}\sqrt{T_{tr}/(T_{tr} + R t)}$ for $0 \leq t \leq t_q$ (or $T_q \leq T \leq T_{tr}$) and $\xi(t) = \xi(T_{exp})$ when $t > t_q$. It is assumed that at $t = t_q$ the temperature has a discontinuity while the domain size is continuous.

If $t^* > t_q$, during cooling, the domains are in equilibrium. The domain size is

$$r(t) = \xi(T(t)) \qquad \text{for } 0 < t < t_q \ ,$$

$$r(t) = \min\left\{r_q + a\,(T_{exp}/T_\sigma)\ln\left[1 + (t - t_q)/\tau(r_q, T_{exp})\right], \xi(T_{exp})\right\} \qquad \text{for } t > t_q \ , \quad (7)$$

with $r_q = \xi(T_q)$.

If $0 < t^* < t_q$, the domains are in equilibrium up to $t^*$ and fall out of equilibrium when $t > t^*$. Afterwards, they non isothermally grow from $r(t^*) = \xi\left(T_{tr} + R\, t^*\right)$ down to $T_q$ and then start to isothermally grow at $T_{exp}$. Therefore, the domain size is

$$r(t) = \xi(T(t)) \qquad \text{for } 0 < t < t^* \;,$$

$$r(t) = r(t^*) + \int_{t^*}^{t} dt'\, \frac{d\rho(t', T(t'))}{dt'} \qquad \text{for } t^* < t < t_q \;, \quad (8)$$

$$r(t) = \min\left\{ r_q + a\, (T_{exp}/T_\sigma) \ln\left[1 + (t - t_q)/\tau(r_q, T_{exp})\right],\, \xi(T_{exp}) \right\} \qquad \text{for } t > t_q \;,$$

with $r_q = \xi(T_q)$.

For $T_q > T^*$, the amplitude of aging of the domain size is

$$\xi(T_{exp}) - \xi(T_q) = \xi_{tr}\, \sqrt{T_{tr}} \left( \frac{1}{\sqrt{T_{exp}}} - \frac{1}{\sqrt{T_q}} \right) \;.$$

For $T^* > T_q$ and if $T^*$ is not far from $T_q$, the same expansion as above reads

$$\xi(T_{exp}) - \rho(t_q) = \xi_{tr}\, \sqrt{T_{tr}} \left( \frac{1}{\sqrt{T_{exp}}} - \frac{1}{\sqrt{T_q}} \right) + \left(\xi(T_q) - \rho(t_q)\right) \;,$$

with

$$\xi(T_q) - \rho(t_q) = \xi_{tr}\, \frac{\left[T^* - T_q\right]^2}{8\, T_{tr}^{1/3}\, T_\sigma^{5/3} \left(\frac{\xi_{tr}}{a}\right)^{5/3}} \left(\ln(R_0/|R|)\right)^{5/3} \left[5 + 3\ln(R_0/|R|)\right] \;. \quad (9)$$

Consequently, for a given cooling rate $R$, there is a regime change at $T_q = T^*$.

We emphasize that for the preceding two calculations the natural energy scale is not the microscopic barrier height $T_\sigma$ but the effective, much larger, barrier height $\tilde{T}_\sigma = T_\sigma\, (\xi_{tr}/a)$.

Temperature cycling experiments are more complex. The comparison of the experiments with the predictions is only performed using numerical computation (see below).

**3.2. Deeper Insight into the Model** - The previous simplified form of the model is able to describe some features actually observed in our experiments. However, it is not fully satisfying for, at least, three reasons. First, the pictures of figure 5 clearly show that there is a distribution of domain sizes. This is not surprising if we know that the elementary growing process of a domain is a dipole flip: if a domain grows, one of its neighbors slims. Second, the effective barrier height $\tilde{T}_\sigma(r_1) = (r_1/a)\, T_\sigma$ may be seen as a mean value over all the frontier of a domain which crosses different regions of the sample where the density of pinning points is at random. Therefore, there must be a distribution of the height $\tilde{T}_\sigma$. A third important point is that without a distribution of barrier heights, it is impossible to reproduce both the isothermal aging and the cooling rate dependence: either the (unique) barrier height is too large, and then there is cooling rate dependence but no aging at the final temperature, or the barrier height is too small and there is barely any cooling rate effects. Therefore, cooling rate effects must be attributed to large local barrier heights, which coexist with smaller barriers. Finally, our recent results on the response to a positive temperature jump from $T_1$ to $T_2$ (see sub-section 2-5) give again a very strong indication in favour of the coexistence of domains with very different sizes. Indeed, the non monotonic variation of $\varepsilon'$ cannot be found, even with a distribution of the domain sizes, if they all grow. The solution is to assume, immediately after the positive jump, that some domains grow because they are still smaller than $\xi(T_2)$ coexisting with slimming domains because they have grown beyond $\xi(T_2)$.

The last point deserves some examination. We have seen that the delayed response of $\varepsilon'$ to a positive temperature jump is non monotonic and composed of a fast increase followed by a slow decrease. One might think that the increase of $\varepsilon'$ would be due to the decrease of domains by dipole flips at the frontier. This would be the inverse process of the growing process when the domains are large and consequently when the process is logarithmically slow. It would provide a very slow slimming process, in contradiction with the experiments. On the contrary, a fast process is provided by the birth (or nucleation) of new very small domains or nuclei inside the too large domains. These nuclei rapidly grow and finally induce the fragmentation of a large domain into new domains with their size equal to the coherence length $\xi(T_2)$. We first examine the fragmentation of a domain with a given size and afterwards we take into account the size distribution.

**3.2.1. Fragmentation of a single domain** - The process begins with the creation of a germ or nucleus which may appear anywhere in the too large domain. Its initial size is of the order of the lattice parameter $a$. It grows as

$$\rho_n(t) = a + a\,(T_2/T_\sigma)\ln[1 + t/\tau(a,T_2)] \quad,$$

but later its real size is

$$r_n(t) = \min\left\{a + a\,(T_2/T_\sigma)\ln[1 + t/\tau(a,T_2)], \xi(T_2)\right\} \quad. \tag{10}$$

The index $n$ stands for nucleus. For the present calculation we take as time origin $t = 0$ the instant of the temperature jump. Some orders of magnitude have to be recalled. The size $r(0)$ of a too large domain is such that $\xi(T_2) \leq r(0) \leq \xi(T_1)$. Our experiments are performed with jumps from $T_1 = 5$ K to $T_2 \leq 20$ K and consequently $1 \leq \xi(T_1)/\xi(T_2) \leq 2$. Therefore, a small number of nuclei growing up to $\xi(T_2)$ are sufficient to break down the large domain into new ones with $r_n(t) \leq \xi(T_2)$.

Let be $\Gamma(T_2)$ the number of nuclei created per volume unit and time unit at the temperature $T_2$. If we take into account that new nuclei are continuously created, the contribution to the wall area built inside a given large domain, is, at short times,

$$\sigma_N(t) \approx \Gamma(T_2)\,r(0)^3 \int_0^t \mathrm{d}t'\,(r_n(t'))^2 \quad, \tag{11}$$

where the index $N$ stands for the whole set of nuclei inside the large domain of size $r(0)$. We note that there is no factor of $1/2$ here because, at least at the beginning of the process, the nuclei are just like islands in a sea, without common coast. This equation is based on two assumptions: i) the size $r(0)$ of the large domain is practically invariant during the fragmentation process; ii) the nuclei are created anywhere in the large domain at any time. The second assumption is unrealistic since new nuclei can only be born in the part of the large volume not occupied by previously created nuclei. In fact, the nucleation process becomes less efficient as long as the number of nuclei increases.

At the beginning of the growing process $t \ll \tau(a,T_2)$, when the second assumption is still satisfied, then $\rho_n(t) \approx a\left[1+(T_2/T_\sigma)\,t/\tau(a,T_2)\right]$ and the newly created area in a single large domain is in lowest order

$$\sigma_N(t) \cong \Gamma(T_2)\, r(0)^3\, a^2\, t \ . \tag{12}$$

At $t=0$ the new area and its time derivative are

$$\sigma_N(0) = 0 \quad \text{and} \quad \left(\frac{d\sigma_N}{dt}\right)_{t=0} \cong \Gamma(T_2)\, r(0)^3\, a^2 \ . \tag{13}$$

The derivative depends on $T_2$ only through $\Gamma(T_2)$. This is a fast initial increase of the area as we need.

The process stops when the whole volume of the large domain is occupied by grown up nuclei. The condition which defines the time $\theta_M$ would read

$$\Gamma(T_2)\, r(0)^3 \int_0^{\theta_M} dt'\, \left(r_n(t')\right)^3 \approx r(0)^3 \tag{14}$$

if the assumption ii) were true. Indeed, the value obtained in this way for $\theta_M$ is only a lower limit. From the definition of $\theta_M$, we have $\sigma_N(\infty) \approx \sigma_N(\theta_M)$. Since $\sigma_N(0)=0$, the total variation is simply $\Delta\sigma_N = \sigma_N(\theta_M)$.

Several cases have to be considered according to the relative efficiencies of the two processes (nucleation and growth) in action. This may be inferred from the first step of an iteration process which takes into account the remaining volume where new nuclei are allowed to appear. It gives

$$\sigma_N(t) \approx \Gamma(T_2) \int_0^t dt' \left[r(0)^3 - \Gamma(T_2)\, r(0)^3 \int_0^{t'} dt''\, \left(r_n(t'')\right)^3\right] \left(r_n(t')\right)^2 \ .$$

For short times, this equation becomes

$$\sigma_N(t) \approx a^2\, \Gamma(T_2)\, r(0)^3 \left\{ t + \left(\frac{(T_2/T_\sigma)}{\tau(a,T_2)} - \frac{\Gamma(T_2)\, a^3}{2}\right) t^2 + O(t^3) \right\} \ . \tag{15}$$

The initial straight line bends downwards if $\Gamma(T_2) a^3 \tau(a, T_2)(T_\sigma/2 T_2) > 1$ (strong nucleation) or bends upwards if $\Gamma(T_2) a^3 \tau(a, T_2)(T_\sigma/2 T_2) < 1$ (fast growth). The physical meaning of the limit condition

$$\Gamma(T_2) a^3 \tau(a, T_2)(T_\sigma/2 T_2) \cong 1 \qquad (16)$$

is that the number of nuclei created in a volume of size $a$ during the characteristic time $\tau(a, T_2)$ of growth is of the order of unity. Indeed three cases have to be distinguished, separated by the two following limits.

Nucleation is said to be very frequent if a new nucleus appears (in the mean) when the previous nucleus has hardly changed its size. The time $\delta t$ needed for doubling the nucleus size is defined by

$$a \left( 1 + \frac{T_2}{T_\sigma} \frac{\delta t}{\tau(a, T_2)} \right) \cong 2 a \quad \text{or} \quad \delta t \cong \frac{T_\sigma}{T_2} \tau(a, T_2) \ .$$

During this time, at least one nucleus must be born. The limit condition reads

$$\Gamma(T_2) a^3 \tau(a, T_2)(T_\sigma/T_2) \geq 1 \ . \qquad (17)$$

It is calculated at the end of the fragmentation process, when the remaining volume is of the order of $a^3$. Within a factor of 2 the two equations (16) and (17) are identical.

Nucleation is said to be very rare if a new nucleus appears (in the mean) only when the previous nucleus has reached its final size $\xi(T_2)$. The time $\Delta t$ needed for this evolution is defined by

$$a \ln \left( \frac{T_2}{T_\sigma} \frac{\Delta t}{\tau(a, T_2)} \right) \cong \xi(T_2) \quad \text{or} \quad \Delta t \cong \tau(a, T_2) \exp\left( \frac{\xi(T_2)}{a} \frac{T_\sigma}{T_2} \right) \ .$$

During this time, at most one nucleus must be born. The limit condition reads

$$\Gamma(T_2) r(0)^3 \tau(a, T_2) \exp\left( \frac{\xi(T_2)}{a} \frac{T_\sigma}{T_2} \right) \leq 1 \qquad (18)$$

It is calculated at the beginning of the fragmentation process, when the remaining volume is of the order of $r(0)^3$.

The two previous limits define three nucleation regimes:

Regime 1: rare nucleation

$$\Gamma(T_2)\,\tau(a,T_2) \leq \frac{1}{r(0)^3 \exp\bigl((\xi(T_2)/a)\,(T_\sigma/T_2)\bigr)} \quad ;$$

Regime 2: intermediate nucleation

$$\frac{1}{r(0)^3 \exp\bigl((\xi(T_2)/a)\,(T_\sigma/T_2)\bigr)} \leq \Gamma(T_2)\,\tau(a,T_2) \leq \frac{1}{a^3\,(T_\sigma/T_2)} \quad ;$$

Regime 3: frequent nucleation

$$\frac{1}{a^3\,(T_\sigma/T_2)} \leq \Gamma(T_2)\,\tau(a,T_2) \quad .$$

The three regimes are examined in detail in Appendix A.

**3.2.2. Distribution of the domain size** - In an experiment with a positive temperature jump to $T_2$ after a sojourn of duration $t_1$ at $T_1$, the volumic fractions of large domains ($\xi(T_2) \leq r(0) \leq \xi(T_1)$) and of small domains ($r(0) \leq \xi(T_2)$) are $P_{la}$ and $P_{sm} = 1 - P_{la}$, respectively. The fraction $P_{la}(t_1, T_2)$ is an increasing function of $T_2$ since when the temperature $T_2$ increases the coherence length $\xi(T_2)$ decreases and consequently, the interval $\xi(T_2) \leq r(0) \leq \xi(T_1)$ becomes broader. It is also an increasing function of $t_1$ which saturates for very long $t_1$ when nearly all the sample is occupied by domains larger than $\xi(T_2)$. Moreover, its dependence on $T_2$ is weakened when $t_1$ is long because most of the domains have size close to $\xi(T_1)$ while only a few of them are still in the size range of $\xi(T_2)$.

Now we have to take into account the size distribution. Two cases must be distinguished: a) a simple experiment (cooling followed by isothermal aging) where all the domains play the same role; b) a more complex experiment (cooling, isothermal aging at $T_1$, positive temperature jump at $T_2$ and finally isothermal aging at this temperature) where two types of domains must be separated: those larger than $\xi(T_2)$ and those smaller than $\xi(T_2)$. In the first case, any mean value is calculated through a sum over all the domains with the convenient weighting function. It is noted by brackets $\langle F \rangle_{dom}$ with the index $dom$. In the second case, we have to add the contribution of the newly created walls through nucleation, and the final average reads: $\langle F \rangle = \langle F \rangle_{dom} + P_{la} \langle F \rangle_N$ where $\langle F \rangle_N$ concerns only the walls of the nuclei.

Now we compare the results predicted by the model with our data. The second contribution in the above formula is well fitted by an exponential return $\Delta\varepsilon_{incr}\left[1-\exp(-t/\tau_{incr})\right]$ (see sub-section 2.5) which has to be compared with the above term $P_{la}\langle F\rangle_N$ which is proportional to $P_{la}\langle\sigma_N\rangle_{la}$.

We first compare our experimental data on the amplitude $\Delta\varepsilon_{incr}$ of the increase of $\varepsilon$ with the prediction $\Delta\varepsilon_N \propto P_{la}\langle\Delta\sigma_N\rangle_{la}$ provided by our model. The dependence on $t_1$ is obviously satisfied: the amplitude is an increasing function of $t_1$ which tends towards an asymptote when $t_1$ becomes very long. Concerning the dependence on $T_2$, things are less clear. On the one hand, the observed amplitude $\Delta\varepsilon_{incr}$ decreases when $T_2$ increases (see figure 3). On the other hand, in the expression for $\Delta\varepsilon$ the factor $P_{la}$ increases with $T_2$. Therefore, in order to have a decreasing product the factor $\langle\Delta\sigma_N\rangle_{la}$ must be decreasing too. This constraint selects the intermediate case and it additionally shows that the increase of the first factor is slower than the decrease of the second one. In this case, the amplitude $\Delta\varepsilon_N$ is

$$\Delta\varepsilon_N \propto P_{la}\left\langle \frac{r(0)^3\, T_\sigma}{2\,a\left[T_2\,\ln(\theta_2/\tau_\infty)-T_\sigma\right]}\right\rangle_{la} \qquad (19)$$

where $\theta_2$ (defined as the time when the volume of a large domain is completely filled with newly created domains) is now given by the mean value

$$\theta_2 \approx \frac{1}{\Gamma(T_2)\,a^3\left\langle\left[(T_2/T_\sigma)\ln(\theta_2/\tau_\infty)-1\right]^3\right\rangle_{la}} \qquad (20)$$

Provided that the dependence of $\Gamma(T_2)$ on temperature is less than Arrhenius, one indeed finds that the second factor decreases as the temperature increases. Finally, we notice that the observed weakening of the $T_2$ dependence when $t_1$ becomes very long is also predicted (from the behavior of $P_{la}$).

Our second comparison is about the characteristic time. If we do the approximation

$$\left\langle\frac{t}{\tau_2}\right\rangle \approx \frac{P_{la}\langle\sigma_N(t)\rangle_{la}}{P_{la}\langle\Delta\sigma_N\rangle_{la}} = \frac{\langle\sigma_N(t)\rangle_{la}}{\langle\Delta\sigma_N\rangle_{la}} \approx \left\langle\frac{\sigma_N(t)}{\Delta\sigma_N}\right\rangle_{la} \qquad \text{for} \quad t \ll \tau_2 \;,$$

the factor $P_{la}$ cancels and we get

$$\langle \tau_2 \rangle \approx \frac{1}{2\,\Gamma(T_2)\,a^3\,\left\langle\left[(T_2/T_\sigma)\ln(\theta_2/\tau_\infty)-1\right]\right\rangle_{la}} \quad . \tag{21}$$

Clearly, the two characteristic times $\tau_2$ and $\theta_2$ both decrease when $T_2$ increases and they do not depend on $t_1$. We recall that they are not equal because they correspond to different situations: the former is relative to the initial slope while the latter concerns the end of the nucleation process.

Our experimental data are well fitted with the exponential law $\Delta\varepsilon_{incr}\left[1-\exp(-t/\tau_{incr})\right]$. This means that the initial variation is governed by the time $\tau_{incr}$ which is found to be nearly independent of $t_1$ and to decrease when the temperature $T_2$ increases (see figure 4). We notice that the comparison with the characteristic times $\tau_1$, $\tau_2$ and $\tau_3$ of the three regimes (defined in Appendix A) is not a criterion of choice. Indeed, it does not rule out any regime since the three times vary like $\tau_{incr}$ (however, with the improbable additional condition for regime 3 that $\Gamma(T_2)\,T_2^{-3/2}$ decreases when temperature increases).

As a conclusion, we emphasize that the main features of the non monotonic behavior which immediately follows a positive temperature jump are quite well accounted for by our modified (and improved) model.

**3.3. Numerical Calculations on the Model** − The different behaviors of the domain size $r(t)$ predicted by our model in its simplified form were computed for a unique barrier height and consequently no size dispersion. We recall that in this case the variation of the dielectric constant is simply proportional to the inverse of the unique domain size $K/r(t)$ and its lowest limit is $K/\xi(T_{exp})$.

The results concerning $r(t)$ are sketched in figure 6 which shows the reduced domain size $r(t)/\xi(T_{exp})$ (full line) for the barrier height $T_\sigma$ and a cooling from temperature $T_{tr}$ to temperature $T_{exp}$ lasting the time $t_c = (T_{tr}-T_{exp})/|R|$ followed by an isothermal evolution. The data are displayed together with a reference. The ascending curve $\xi(T(t))/\xi(T_{exp}) = \left(\xi(T_{tr})/\xi(T_{exp})\right)\sqrt{T_{tr}/(T_{tr}+R\,t)}$ shows the upper limit for $r(t)/\xi(T_{exp})$ (dashed line). If $T_\sigma$ is large, the size cannot increase and $r(t)=\xi_{tr}$ at any time. If $T_\sigma$ is very small, the domains are always in equilibrium and $r(t)=\xi(T(t))$ until $t=t_c$ and then $r(t) = \xi(T_{exp})$. For intermediate values of $T_\sigma$ such as that used in figure 6, the representative point also begins with a travel on the curve corresponding to $r(t) = \xi(T(t))$ until $t = t^*$ (or $T = T^*$) where it leaves the curve, then goes slowly upwards and after $t = t_c$, logarithmically increases. Two different time scales are used in the two half-figures in order to emphasize these features: a) the left half-figure covers the cooling time $t_c$ and the splitting between $r(t)$ and $\xi(T(t))$ is clearly seen at time $t^*$ or temperature $T^*$; b) the right half-figure covers $10^5 \times t_c$ and the very slow isothermal growth of $r(t)/\xi(T_{exp})$ becomes obvious. This illustrates very well the set of equations (3).

Indeed, from the experimentalist's point of view, it is more convenient to take as time origin the instant when isothermal aging begins. Hence, the relevant time is either $\hat{t} = t - t_c$ (full ramp) or $\hat{t} = t - t_q$ (quenching). In the experiments, the variable $t$ is indeed $\hat{t}$. In what follows we adopt this new time origin. Here we present the results of our computations on the time dependent part of the dielectric constant (in arbitrary units) during the isothermal evolution following a temperature ramp in the form $\delta\varepsilon(\hat{t}) = K\left(1/r(\hat{t}) - 1/\xi(T_{exp})\right)$. In this way, if the domains have reached their equilibrium size $\xi(T_{exp})$, then $\delta\varepsilon = 0$. Our data are displayed in Table I. We have calculated the values of $\delta\varepsilon(\hat{t})$ at $\hat{t} = 0$ s (beginning of the isothermal evolution) and $\hat{t} = 10^8$ s (a duration much longer than our experiments) for different values of the unique effective barrier height $\tilde{T}_\sigma = T_\sigma\,(\xi_{tr}/a)$. We have used for the cooling rate $R$ two extreme values which span the range of our experimental values.

**<u>Table I</u>**

These results deserve some comments. For $\tilde{T}_\sigma = 40$ K the domains always have their equilibrium size: there is no aging. For $\tilde{T}_\sigma = 60$ K the domains have not their equilibrium size at the end of the ramp, but they are able to reach it in less than $10^8$ s: this is ergodic aging. For $\tilde{T}_\sigma = 90$ K and $\tilde{T}_\sigma = 100$ K the domains have not their equilibrium size at the end of the ramp and they are not able to reach it within $10^8$ s: this is aging with effective ergodicity breaking (effective, since the model is fundamentally ergodic). For $\tilde{T}_\sigma = 150$ K the domains have not their equilibrium size at the end of the ramp and the growing process is so slow that it cannot be observed during $10^8$ s: this looks like the case of no aging, except for the dependence on *R*.

Two other features can be seen in Table I. Firstly, the maximum possible evolution $\delta\varepsilon_{max} = \delta\varepsilon(0) - \delta\varepsilon(\infty) = \delta\varepsilon(0)$ monotonously increases with $\tilde{T}_\sigma$. Secondly, the evolution observed during a long time $\delta\varepsilon_{long} = \delta\varepsilon(0) - \delta\varepsilon(10^8\text{s})$ is equal to 0 for both low barriers and high barriers, and passes through a maximum for $\tilde{T}_\sigma$ in the middle range. This is easy to explain. For small $\tilde{T}_\sigma$ the domains rapidly grow, they are able to reach the coherence length within a reasonable time, $\delta\varepsilon_{long}$ is equal to $\delta\varepsilon_{max}$, and they both tend to 0 if $\tilde{T}_\sigma$ is very small. For large $\tilde{T}_\sigma$ the domains can hardly grow, during the ramp they shortly leave the coherence length, they cannot reach the equilibrium in a finite time, and $\delta\varepsilon_{long}$ is smaller than $\delta\varepsilon_{max}$. If $\tilde{T}_\sigma$ is very large, they do not grow at all at $T_{exp}$, and therefore $\delta\varepsilon_{long} \cong 0$. For intermediate $\tilde{T}_\sigma$ values, $\delta\varepsilon_{long}$ which starts from 0 and goes to 0, passes through a maximum.

These computations provide other interesting informations. We call $\varepsilon_{min}$ the value obtained after $400\,000$ s. This is the duration of our longest experiments. Then we have shown: i) $\varepsilon_{min}$ is nearly (but not perfectly) a linear function of the logarithm of the cooling rate, at least for $0.001\,\text{K.s}^{-1} \leq |R| \leq 1\,\text{K.s}^{-1}$; ii) the cooling rate *R* and the relaxation time $\tau_\infty$ exactly play the same role (indeed, $\varepsilon_{min}$ is a function of the product $R\,\tau_\infty$ only).

Numerical calculations were also performed when there is a sudden change from temperature $T_q$ to temperature $T_{exp}$. The dependence of $\varepsilon_{min}$ on the quenching temperature $T_q$ is strong for $T_q$ close to $T_{tr}$ and weak at low temperatures. Temperature cycles were also simulated.

Finally, we have seen that the simplified form of our model provides a rather good qualitative agreement with our data for most of our experiments: simple aging, quenching, positive or negative temperature cycles. However, as discussed above, some qualitative features (such as the coexistence of cooling rate effects and isothermal aging) require some distribution of the domain size. We try now to deduce more information on this distribution, and find interesting qualitative conclusions.

## 4. Distribution of barrier heights

In a series of experiments all performed with the same cooling rate $R = -0.015 \text{ K.s}^{-1}$ on a KLT sample with the lithium concentration $x = 0.017$, we have observed that the amplitude of the decay $\delta\varepsilon = \delta\varepsilon(0) - \delta\varepsilon(10^4 \text{ s})$ during $10000 \text{ s}$ varies in a non monotonic manner with the temperature $T_{exp}$ of the isothermal aging. The data are displayed as full squares in figure 7. They present a broad maximum near $T = 15 \text{ K}$. It is actually impossible to reproduce this feature with a single barrier height model. Rather, we attempt to determine the full distribution function $P(T_\sigma)$ of the barrier heights using this effect. We choose to parametrize it using a gamma distribution which contains only two free parameters and gives a rather good fit. This normalized function reads

$$P(T_\sigma) = \frac{Y(T_\sigma)}{\Gamma(\nu)\,\Theta} \left(\frac{T_\sigma}{\Theta}\right)^{\nu-1} \exp\left(-\frac{T_\sigma}{\Theta}\right) \quad \text{with} \quad \nu > 0 , \quad (23)$$

where $Y(T_\sigma)$ is the Heaviside step function and $\Gamma(\nu)$ is the gamma function. The distribution is scaled by the parameter $\Theta$, related to the most probable barrier by $T_\sigma^{mp} = (\nu - 1)\,\Theta$ if $\nu > 1$; if $0 < \nu \leq 1$ then $T_\sigma^{mp} = 0$.

In figure 7 are shown the data together with the results of the calculation with three values of the exponent $\nu$. For $\nu = 1$ (then $P(T_\sigma)$ is simply a decreasing exponential) the curve steadily decreases, in clear contradiction with the experiments. The curves for the parameter $\Theta = 60 \text{ K}$ and the exponents $\nu = 2.2$ and $\nu = 3$ both present a maximum in the right place. However, for $\nu = 3$ the curvature is too sharp, while for $\nu = 2.2$ the fit to the data is quite good. Moreover, we have established that the position of the maximum of the curve roughly varies as the most probable value $T_\sigma^{mp}$. These results mean that the data cannot be reproduced by a distribution which does not vanish for small barriers. More precisely, a good fit needs that $\nu \cong 2$, which implies that the distribution varies as $P(T_\sigma) \approx T_\sigma$ for $T_\sigma \ll \Theta$, and that $T_\sigma^{mp} \cong 60 \text{ K}$. We now adopt these values. Then the corresponding mean value is $T_\sigma^{mean} \cong 100 \text{ K}$ and the full width at half height is $\Delta T_\sigma \cong 155 \text{ K}$. In principle, the result is valid only for the $x = 0.017$ sample, but we assume that it retains the characteristics of all our KLT samples. We therefore use the distribution

$$P(T_\sigma) = Y(T_\sigma) \frac{T_\sigma}{\Theta^2} \exp\left(-\frac{T_\sigma}{\Theta}\right) . \tag{23}$$

The other needed numerical quantities are the lattice parameter $a = 4 \times 10^{-10}$ m, the coherence length at the transition temperature $\xi(T_{tr}) = 5\,a$ and the relaxation time $\tau_\infty = 10^{-10}$ s. The computed quantity is simply the ratio of mean values $\langle \rho^2 \rangle / \langle \rho^3 \rangle$ the variation of which is proportional to $\delta\varepsilon(t)$, normalized at the largest value.

We only report our numerical calculations for the very important results concerning the dependence of the asymptotic value of $\varepsilon(t)$ as the limit of the decay following a temperature ramp at the constant cooling rate $R$. The experiments clearly show a logarithmic dependence [8], also predicted by the model. The results from the model are displayed in figure 8. They show that the asymptotic value depends on the cooling rate as $\varepsilon = p + q \log|R|$ at least in the range from $-0.001\,\text{K.s}^{-1}$ to $-0.1\,\text{K.s}^{-1}$, as in our experiments. The logarithmic variation is perfectly obeyed, actually more accurately than it is when the curve is calculated without distribution of the barrier height. This is again very satisfactory.

Moreover, we have examined the role of the barrier distribution on the coefficient $q$ of the logarithm. We have found that $q$ depends on the unique parameter $T_\sigma^{mp}$ of the distribution in a non monotonic manner: the coefficient $q(T_\sigma^{mp})$ passes through a maximum for $T_\sigma^{mp} \cong 110$ K. Our experiments [8] show that $q$ increases with the lithium concentration $x$ when it is smaller than $x_c \cong 0.02$ and the Li$^+$ ions freeze at $T_{tr}$ into the dipolar glass phase. Indeed, when $x$ increases, the density of pinning points in the crystal increases too, leading to larger equivalent barriers and consequently to a larger $T_\sigma^{mp}$. This is in agreement with our computations. When the concentration $x$ is larger than $x_c \cong 0.02$ the Li$^+$ ions undergo a transition at $T_{tr}$ towards the ferroelectric phase and we observe that the coefficient $q$ then takes small values [8]. We attribute this decrease not to a low $T_\sigma^{mp}$ value but on the contrary to a large one, larger than $T_\sigma^{mp} \cong 110$ K. This is probably due to the correlations between the random fields which must exist in this phase. Then the large density of correlated pinning points is equivalent to a barrier distribution shifted towards high energies.

## 5. Conclusion

In this paper we present an extended version of a brief report [10] on the model elaborated in order to explain a large series of experiments performed on a family of disordered dielectric crystals. The central idea of the model is to attribute the slow isothermal decay of the alternative dielectric constant $\varepsilon(\omega,t)$ to the motion of the walls of the polarized regions or domains of the crystal lattice. This motion is extremely slow because it is hindered by the static random fields associated with the lithium ions which are frozen at the experiment temperatures; moreover it becomes slower and slower as long as its size increases. Thus the existence of such long-lived polarization regions in the paraelectric phase allows us to use the words domain and wall.

In addition to well known features of slow dynamics and aging, such as effective ergodicity breaking with logarithmic dependence on the cooling rate, we report new experiments on a rather surprising non monotonic response of $\varepsilon$ to positive temperature jumps.

Along the lines of the model we calculate the evolution of the typical domain size, assuming in a first step that the growth is slowed down by the same barrier height for all the domains. Even with this oversimplified assumption we are able to qualitatively account for many of the observed effects.

However, the very coexistence of cooling rate effects and isothermal aging, the non monotonic time dependent response to positive temperature jumps and the non monotonic temperature dependence of the amplitude of isothermal decay can be explained only if a broad distribution of the domain barrier height is taken into account. It is satisfying that these three apparently uncorrelated features lead to the same conclusion. Therefore, in a second step we have computed some typical behaviors taking into account the distribution of barrier height $T_\sigma$; we have shown that it goes to 0 with $T_\sigma$ and, consequently, that it has a maximum for a non zero value of $T_\sigma$.

As a matter of fact, if a broad barrier distribution is necessary to explain the non monotonic response to a temperature jump, it is not sufficient to induce the observed initial rapid variation. For that purpose, we have introduced a nucleation process which is known to be fast because it concerns small entities. The results are then in good qualitative agreement with experiments.

Finally, we emphasize that the model is not very complex since the dynamics of the domains is completely characterized by a single time dependent length $r$ and the function which describes the barrier distribution, itself parametrized by its most probable value. Nevertheless, this model is quite powerful since it describes semi-quantitatively all the features observed in KLT crystals.

# APPENDIX A

The three different nucleation regimes are now considered, taking into account the decay of the remaining volume in the too large domain.

<u>Regime 1</u> - If nucleation is very strong while growth is slow, the process stops at $\theta_M = \theta_1$ when the total volume of the large domain is filled by very small domains of typical size $a$ which had not enough time to grow. Under this oversimplified hypothesis, the unrealistic assumption that the nuclei are created anywhere in the large domain at any time can be discarded. Then the evolution of the number $N(t)$ of nuclei is governed by the equation

$$\frac{dN}{dt} = \Gamma(T_2)\, a^3 \left( \frac{r(0)^3}{a^3} - N \right) ,$$

the solution of which is

$$N(t) = \frac{r(0)^3}{a^3} \left( 1 - \exp(-\Gamma(T_2)\, a^3\, t) \right) .$$

Consequently, the total nucleus area is

$$\sigma_N(t) = \frac{r(0)^3}{a} \left( 1 - \exp(-\Gamma(T_2)\, a^3\, t) \right) . \tag{A-1}$$

The expansion of equation (A-1) for short times leads to

$$\sigma_N(t) = \Gamma(T_2)\, r(0)^3\, a^2 \left( t - \frac{\Gamma(T_2)\, a^3}{2} t^2 + O(t^3) \right) .$$

As expected, it is identical to equation (15) with $\tau(a, T_2) = \infty$. With $\tau_1 = 1/(\Gamma(T_2)\, a^3)$, the fragmentation process is practically achieved for $\theta_1 \approx 10\, \tau_1$.

Indeed, the final total area is $\sigma_N(\infty) \approx r(0)^3/(2\, a)$ since a factor of $1/2$ has to be included since all the domains are in close contact and consequently the total variation is $\Delta\sigma_N \approx r(0)^3/(2\, a)$. Therefore, we adopt as a useful approximation the equation

$$\sigma_N(t) = \frac{r(0)^3}{a}\, \varphi(t) \left( 1 - \exp(-\Gamma(T_2)\, a^3\, t) \right) . \tag{A-2}$$

The slowly decreasing function $\varphi(t)$ varies from 1 for $t = 0$ (isolated nuclei) to $1/2$ for $t \to \infty$ (nuclei in close contact). We notice that $\Delta\sigma_N$ does not depend on $T_2$.

In this case, the new domains have still to grow towards the equilibrium size $\xi(T_2)$ for $t > \theta_1$. This is the standard aging process which induces a decrease of the dielectric constant.

<u>Regime 3</u> - If growth is very fast and nucleation is very slow, when a nucleus was born at time $t'$, its size is $r_n(t) = \min\left\{a + a\,(T_2/T_\sigma)\ln\left[1 + (t-t')/\tau(a,T_2)\right], \xi(T_2)\right\}$ at time $t \geq t'$ and it reaches its final size $\xi(T_2)$ after the delay $\Delta t = t - t' \approx \tau(a,T_2)\exp\left[(\xi(T_2)/a)(T_\sigma/T_2)\right]$. The process stops at $\theta_M = \theta_3$ when the total volume of the large domain is filled by domains of typical size $\xi(T_2)$. Then the evolution is exactly the same as in regime 1 with $a$ changed into $\xi(T_2)$. Here the average number of such new domains in their host domain is equal to $(r(0)/\xi(T_2))^3$. Consequently, the final total area is $\sigma_N(\infty) = r(0)^3/(2\,\xi(T_2))$. The total variation $\Delta\sigma_N = r(0)^3/(2\,\xi(T_2))$ is an increasing function of $T_2$ and the area evolution is governed by the equation

$$\sigma_N(t) = \frac{r(0)^3}{\xi(T_2)} \varphi(t) \left(1 - \exp(-\Gamma(T_2)\,\xi(T_2)^3\,t)\right) \ . \tag{A-3}$$

With $\tau_3 = 1/(\Gamma(T_2)\,\xi(T_2)^3)$, the fragmentation process is practically achieved for $\theta_3 \approx 10\,\tau_3$. In this case the new domains have reached the equilibrium size $\xi(T_2)$ and they no longer evolve after $\theta_3$.

<u>Regime 2</u> - In the two previous regimes when a new nucleus appears the already present nuclei are not evolving: in regime 1 because growth is extremely slow; in regime 3 because growth is achieved. On the contrary, in the intermediate case when a new nucleus was born the present nuclei are still growing; indeed most of them have not yet attained the equilibrium size $\xi(T_2)$ and they grow further.

If the assumption that the nuclei are created anywhere in the large domain at any time is maintained the characteristic time $\theta_M = \theta_2$ is determined by the equation

$$\Gamma(T_2)\,a^3 \int_0^{\theta_2} dt' \left((T_2/T_\sigma)\ln\left[t'/\tau(a,T_2)\right]\right)^3 \approx 1 \ .$$

As already pointed out, the logarithm is a slowly varying function. Consequently, this function may be taken as constant in the integral, with $t' \approx \theta_2$. This approximation, together with $\tau(a,T_2) = \tau_\infty \exp(T_\sigma/T_2)$, leads to

$$\Gamma(T_2)\, a^3\, \theta_2 \left[(T_2/T_\sigma) \ln(\theta_2/\tau_\infty) - 1\right]^3 \approx 1 \ . \tag{A-4}$$

As usual, the value of $\theta_2$ in the logarithm is not important and equation (A-4) provides a lower limit for $\theta_2$ which reads

$$\theta_2 \approx \frac{T_\sigma^3}{\Gamma(T_2)\, a^3 \left[T_2 \ln(\theta_2/\tau_\infty) - T_\sigma\right]^3} \ . \tag{A-5}$$

The creation of nuclei is more efficient if $T_2$ is higher since the rate $\Gamma(T_2)$ is an increasing function of $T_2$. Moreover, it is independent of the enclosing domain size $r(0)$. Therefore, equation (A-5) shows that $\theta_2$ becomes shorter when $T_2$ is increased and that it depends neither on the time $t_1$ elapsed at $T_1$ nor on the size $r(0)$ of the large domain. In the present case the average number of such new domains in their host domain is equal to $\left[r(0)/((T_2/T_\sigma)\, a \ln[\theta_2/\tau(a,T_2)])\right]^3$. Consequently, the final total area is $\sigma_N(\infty) = r(0)^3/\left(2\, (T_2/T_\sigma)\, a \ln[\theta_2/\tau(a,T_2)]\right)$. The total variation is

$$\Delta \sigma_N = \frac{r(0)^3}{2\, a \left((T_2/T_\sigma) \ln(\theta_2/\tau_\infty) - 1\right)} \ , \tag{A-6}$$

a decreasing function of $T_2$. The stopping condition implies that $\rho_n(\theta_2) < \xi(T_2)$. Therefore, when $t > \theta_2$ the new domains still have to grow towards the equilibrium size $\xi(T_2)$. This is the standard aging process which induces the decrease of the dielectric constant.

The time variation is expected to be of the form $\sigma_N(t) = \Delta \sigma_N\, f(t)$ with $f(0) = 0$, $f(\infty) = 1$ and $f(t) \propto t$ for small $t$. A convenient function is $f(t) = 1 - \exp(-t/\tau_2)$. Then

$$\left(\frac{d\sigma_N}{dt}\right)_{t=0} = \frac{\Delta \sigma_N}{\tau_2} \ ,$$

and with equations (A-5) and (A-6) it comes

$$\tau_2 = \frac{1}{2\,\Gamma(T_2)\,a^3\left((T_2/T_\sigma)\ln(\theta_2/\tau_\infty)-1\right)}\,. \qquad (A\text{-}7)$$

According to this equation, the characteristic time $\tau_2$ does not depend on the time $t_1$ elapsed at $T_1$ but it may depend on $T_2$.

Our experiments show that the total increase $\Delta\sigma_N$ is smaller when $T_2$ is higher. This feature necessarily eliminates the first two cases but keeps the last one (see sub-section 3.2.2).

# TABLE I

| $\tilde{T}_\sigma$ (K) | $R = -0.001$ K.s$^{-1}$ | | $R = -0.1$ K.s$^{-1}$ | |
|---|---|---|---|---|
| | $\hat{t} = 0$ s | $\hat{t} = 10^8$ s | $\hat{t} = 0$ s | $\hat{t} = 10^8$ s |
| 40 | 0 | 0 | 0 | 0 |
| 60 | 6 | 0 | 17 | 0 |
| 90 | 33 | 26 | 45 | 28 |
| 100 | 41 | 40 | 54 | 49 |
| 150 | 73 | 73 | 88 | 88 |

# TABLE CAPTION

**Table I** - Computation of the time dependent part $\delta\varepsilon(\hat{t}) = K\left(1/r(\hat{t}) - 1/\xi(T_{exp})\right)$ of the dielectric constant at the beginning ($\hat{t} = 0$ s) of the isothermal evolution following a temperature ramp and after a long time ($\hat{t} = 10^8$ s). A single domain size $r(\hat{t})$ is assumed, the growth of which is characterized by the unique effective barrier height $\tilde{T}_\sigma = T_\sigma\,(\xi_{tr}/a)$. Two extreme values of the cooling rate $R$ are used. The data are given in arbitrary units.

# FIGURE CAPTIONS

**Figure 1** - Variations of the capacitance $C$ after positive temperature jumps from $T_1 = 4.3$ K to several higher temperatures $T_2$. From top to bottom: $T_2 = 11.3$ K; $T_2 = 28.9$ K; $T_2 = 33.8$ K; $T_2 = 39.2$ K. The transition temperature of the sample is $T_{tr} = 35$ K, corresponding to $x = 0.017$. The measurement frequency was $f = 100$ kHz. Before the jump, the sojourn at $T_1$ lasted $t_1 = 5000$ s. For clarity the curves are vertically shifted relative to each other.

**Figure 2** - Variations of the capacitance $C$ after positive temperature jumps from $T_1 = 4.3$ K to $T_2 = 16.2$ K after several durations $t_1$ of the sojourn at $T_1$. From top to bottom: $t_1 = 500$ s, $t_1 = 10000$ s and $t_1 = 80000$ s. The transition temperature of the sample is $T_{tr} = 35$ K. The measurement frequency was $f = 100$ kHz. The temperature jump times are taken as the common time origin.

**Figure 3** - Amplitude $\Delta\varepsilon_{incr}$ of the variation $\Delta\varepsilon_{incr}\left[1 - \exp(-t/\tau_{incr})\right]$ of the dielectric constant after positive temperature jumps as a function of the duration $t_1$ of the sojourn at $T_1$. The temperature jumps are from $T_1 = 4.3$ K to $T_2 = 16.2$ K (squares), $T_2 = 18.7$ K (circles) and $T_2 = 23.8$ K (triangles). The transition temperature of the sample is $T_{tr} = 35$ K. The measurement frequency was $f = 100$ kHz. The data are displayed on logarithmic scales. The amplitude is given in arbitrary units.

**Figure 4** - Characteristic time $\tau_{incr}$ of the variation $\Delta\varepsilon_{incr}\left[1 - \exp(-t/\tau_{incr})\right]$ of the dielectric constant after positive temperature jumps as a function of the duration $t_1$ of the sojourn at $T_1$. The temperature jumps are from $T_1 = 4.3$ K to $T_2 = 16.2$ K (squares), $T_2 = 18.7$ K (circles) and $T_2 = 23.8$ K (triangles). The transition temperature of the sample is $T_{tr} = 35$ K. The measurement frequency was $f = 100$ kHz.

**Figure 5** - Monte Carlo simulation of the three-dimensional random field Ising model at constant temperature $T < T_g$. The figures are snapshots of a cross-section of the system, taken at four times $t_1 < t_2 < t_3 < t_4$ after preparation at $t = 0$. The initial condition is random $S_i = \pm 1$ with probability $1/2$ and corresponds to an infinitely rapid quench from $T = \infty$ to $T$ which has been performed at $t = 0$.

**Figure 6** - Reduced domain size $r(t)/\xi(T_{exp})$ (full line) for a unique barrier height $T_\sigma = 18\,\text{K}$ and a cooling from $T_{tr} = 35\,\text{K}$ to $T_{exp} = 5\,\text{K}$ at the rate $R = -0.01\,\text{K.s}^{-1}$ followed by an isothermal evolution. The reduced coherence length $\xi(T(t))/\xi(T_{exp})$ is also shown (dashed line). Using two different time scales, two important features are put in evidence: a) the left half-figure covers $3000\,\text{s}$ and the time $t^*$ where $r(t)$ and $\xi(T(t))$ split off from each other is clearly seen; b) the right half-figure covers $3 \times 10^8\,\text{s}$ and the very slow isothermal growth of $r(t)/\xi(T_{exp})$ becomes obvious.

**Figure 7** - Amplitude of the isothermal decay of the capacitance $C$ (normalized by the largest value) during $10\,000\,\text{s}$, measured at $f = 100\,\text{kHz}$, as a function of the temperature $T_{exp}$ after the same cooling rate $R = -0.015\,\text{K.s}^{-1}$. The data points (full squares) were obtained on the sample with $x = 0.017$. The three curves are calculated with the gamma distribution (see text). The exponents are $\nu = 1.0$ (dotted line), $\nu = 2.2$ (full line) and $\nu = 3$ (dashed line), respectively. The best fit is obtained for $\nu = 2.2$.

**Figure 8** - The dielectric constant (in arbitrary units) calculated with the gamma distribution (exponent $\nu = 2$, most probable value at $60\,\text{K}$) after an isothermal decay lasting $400\,000\,\text{s}$ at $T = 5\,\text{K}$ and displayed as a function of the cooling rate on a semi logarithmic scale. The calculated points exactly fall on a straight line.

# REFERENCES

[1] PALMER R. G., *Adv. Phys.* **31** (1982) 669.

[2] BOUCHAUD J.-P., *J. Phys. I France* **2** (1992) 1705.

[3] HÖCHLI U. T., KNORR K. and LOIDL A., *Adv. Phys.* **39** (1990) 405.

[4] VUGMEISTER B. E. and GLINCHUK M., *Rev. Mod. Phys.* **62** (1990) 993.

[5] KLEEMANN W., SCHÖNKNECHT V. SOMMER D. and RYTZ D., *Phys. Rev. Lett.* **66** (1991) 762.

[6] DOUSSINEAU P., LEVELUT A. and ZIOLKIEWICZ S., *Europhys. Lett.* **33** (1996) 539.

[7] DOUSSINEAU P., LEVELUT A. and ZIOLKIEWICZ S., *Europhys. Lett.* **33** (1996) 391.

[8] ALBERICI F., DOUSSINEAU P., and LEVELUT A., *J. Phys. I France* **7** (1997) 329.

[9] ALBERICI-KIOUS F., DOUSSINEAU P., and LEVELUT A., *Europhys. Lett.* **39** (1997) 329.

[10] ALBERICI-KIOUS F., BOUCHAUD J.-P., CUGLIANDOLO L. F., DOUSSINEAU P. and LEVELUT A., *Phys. Rev. Lett.* **81** (1998) 4987.

[11] COMÉS R. and SHIRANE G., *Phys. Rev. B* **5** (1972) 1886.

[12] SCOTT J. F., *Rev. Mod. Phys.* **46** (1974) 83.

[13] VAN DER KLINK J. J., RYTZ D., BORSA F. and HÖCHLI U. T., *Phys. Rev. B* **27** (1983) 89.

[14] VINCENT É., HAMMANN J., OCIO M., BOUCHAUD J. P., CUGLIANDOLO L. F., in *Sitges Conference on Glassy Materials*, edited by M. Rubi (Springer-Verlag, Berlin, 1997).

[15] HÖCHLI U. T., WEIBEL H. E. and BOATNER L. A., *J. Phys. C* **12** (1979) L563.

[16] STACCHIOTTI M. G. and MIGONI R. L., *J. Phys. Cond. Matter* **2** (1990) 4341.

[17] NATTERMANN T., *Ferroelectrics.* **104** (1990) 171.

figure1

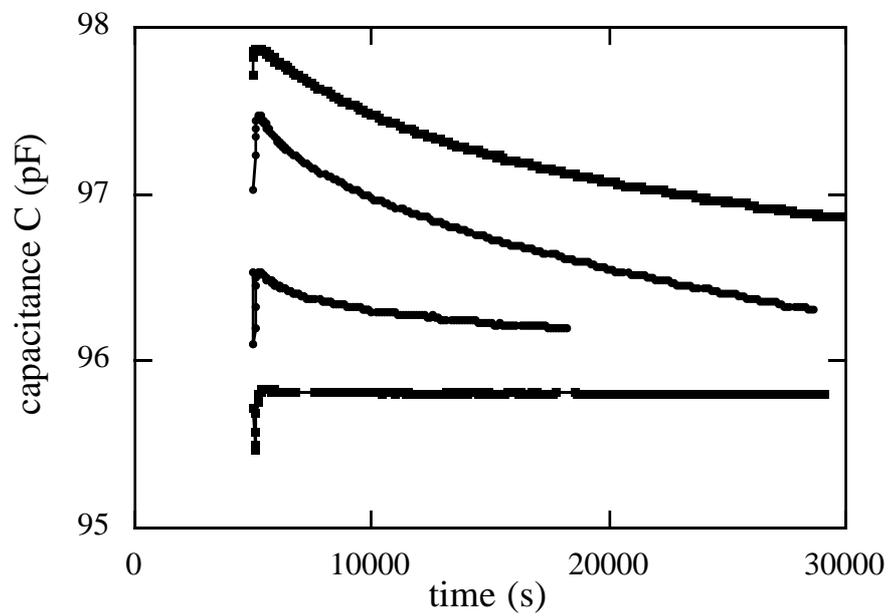

figure 2

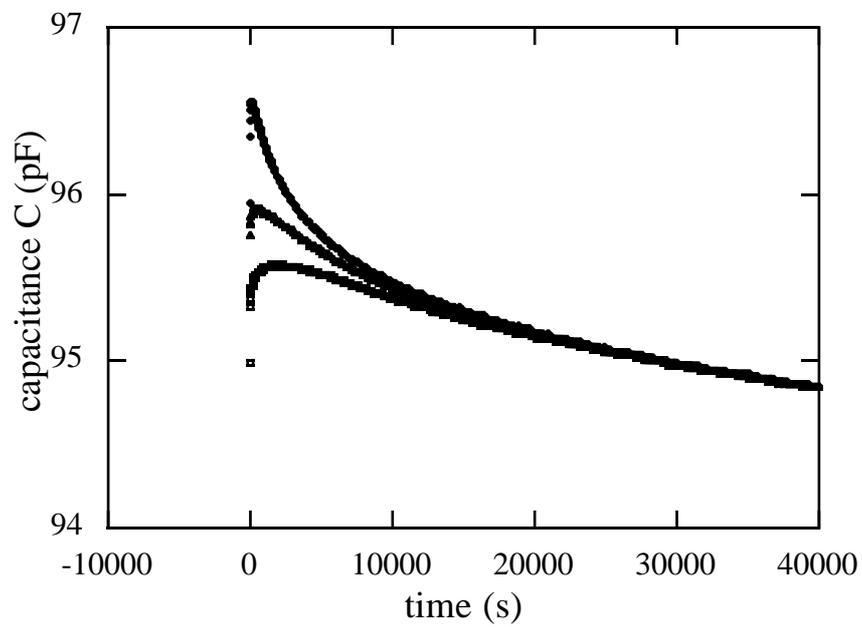

figure 3

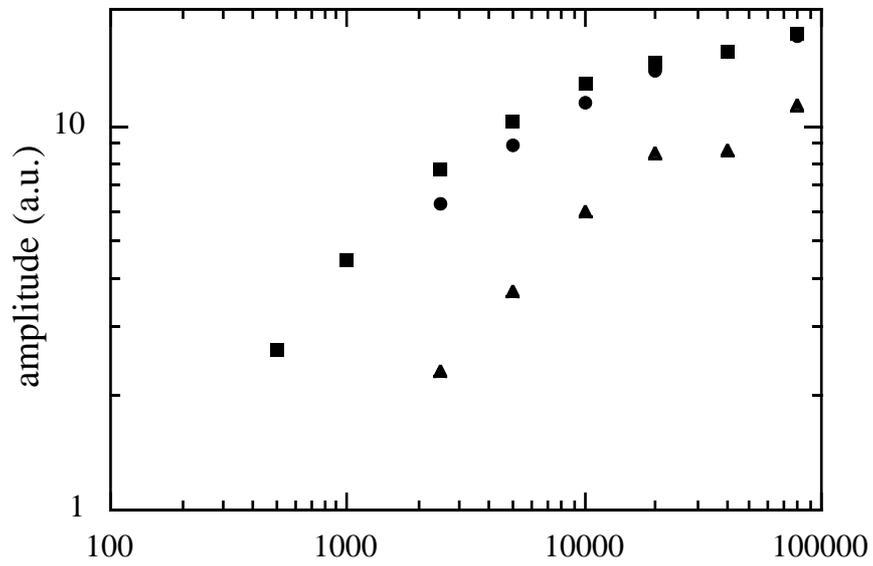

figure 4

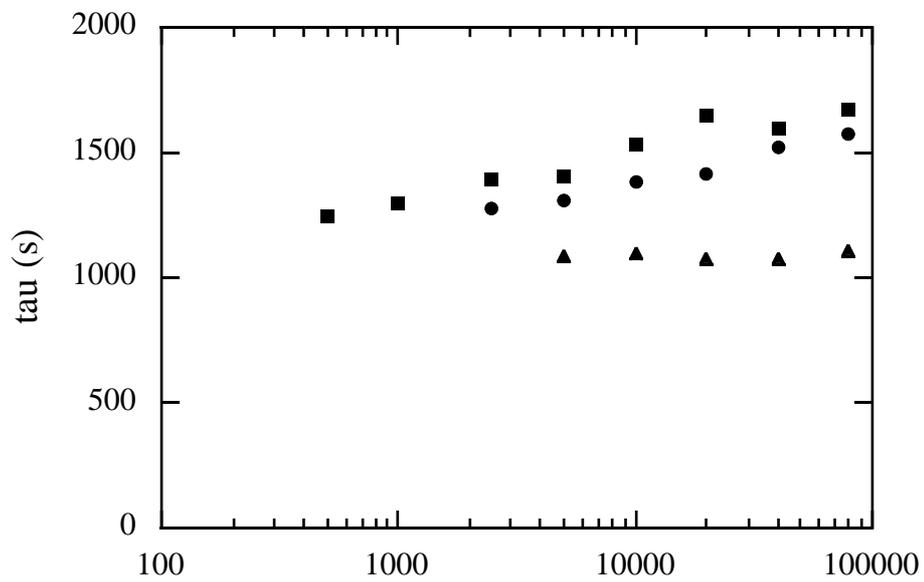

figure 5

**figure 5a**

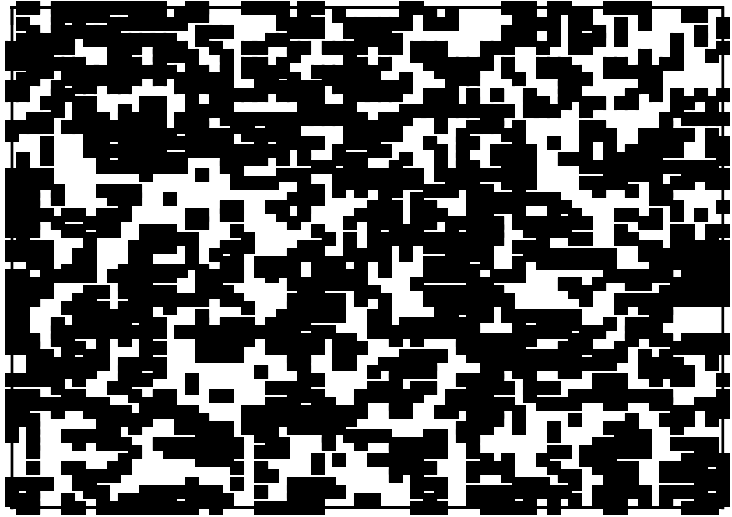

**figure 5b**

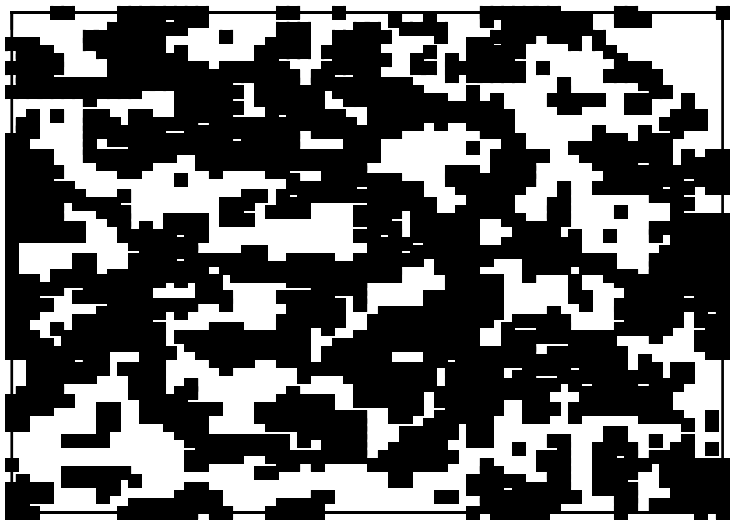

**figure 5c**

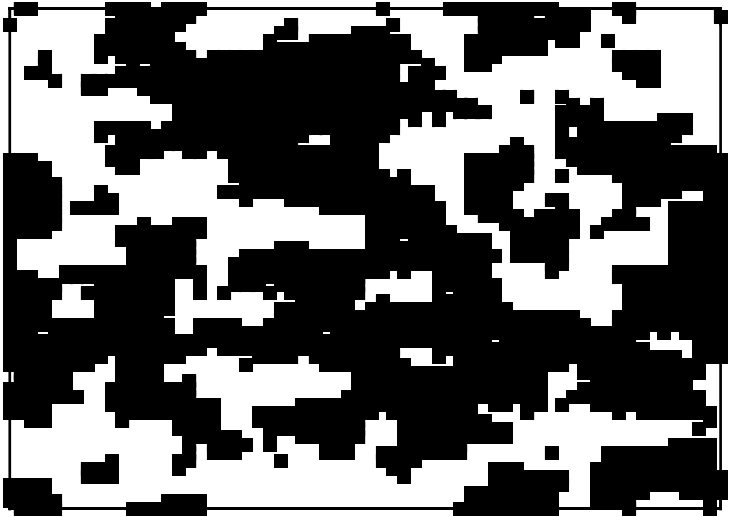

**figure 5d**

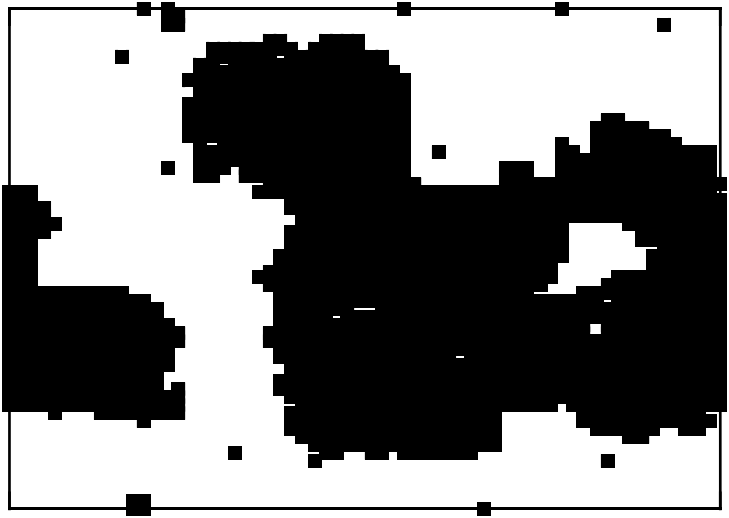

figure 6

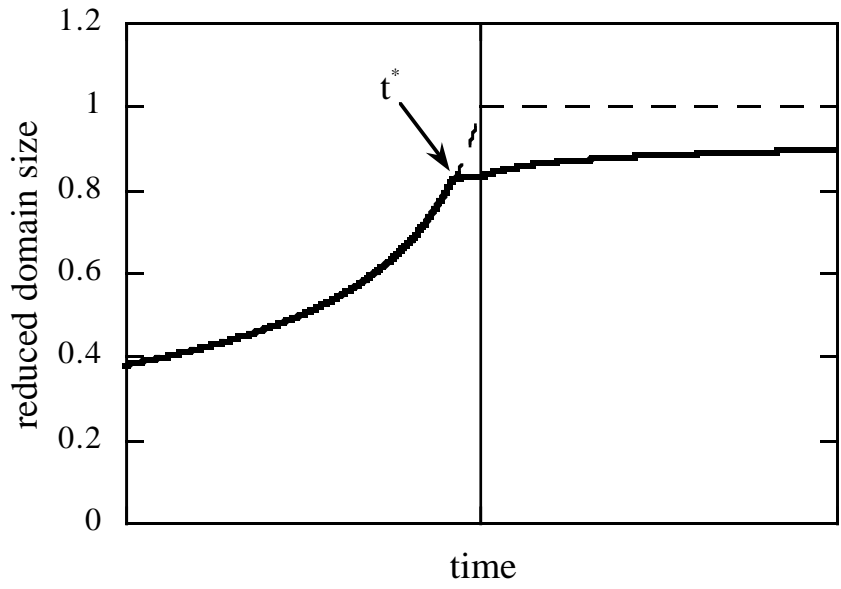

figure 7

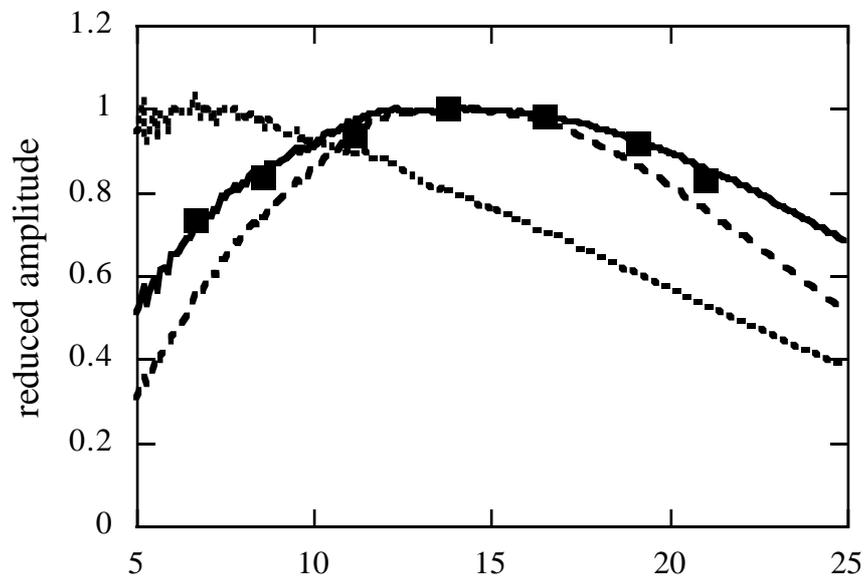

figure 8

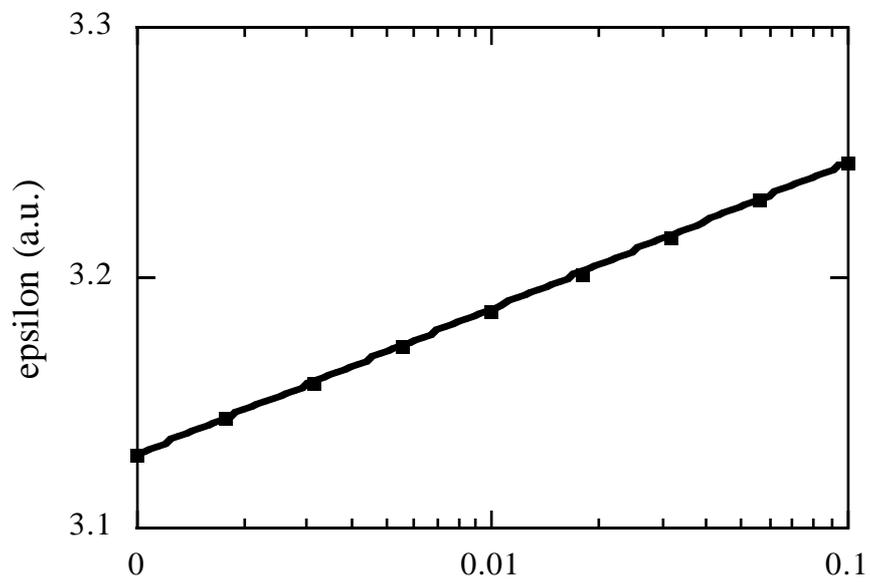